\documentclass[journal,onecolumn]{IEEEtran}
\usepackage{epsfig}
\usepackage{graphicx}
\usepackage{graphics}
\usepackage{epstopdf} 
\usepackage{tikz}
\usetikzlibrary{calc,shapes.geometric}
\usetikzlibrary{arrows,snakes,backgrounds}

\ifCLASSINFOpdf
\else
\fi

\usepackage[cmex10]{amsmath}

\def\clock{\n=\time \divide\n 60
  \m=-\n \multiply\m 60 \advance\m \time
  \ifnum \n>12 \advance\n -12 \fi
   \number\n.\twodigits\m~\ampm\time}
\def\ampm#1{\ifnum #1< 720 am\else pm\fi}
\def\twodigits#1{\ifnum #1<10 0\fi \number#1}

\usepackage{epsfig}

\def\hyptest{\renewcommand{\arraystretch}{-0.7} 
\begin{array}{c}  
\mbox{\tiny{$H_{1}$}}  \\ \vspace{-0.5 mm}
>\\ 
<\\  
\mbox{\tiny{$H_{0}$}} 
\end{array}
}

\def\nexto{\kern -0.54em}

\def\prob{{\rm {I\ \nexto P}}}

\def\pfa{{\rm P_{FA}}}
\def\pd{{\rm P_{D}}}

\def\A{{\cal A}}
\def\B{{\cal B}}

\newcount\m \newcount\n
\def\clock{\n=\time \divide\n 60
  \m=-\n \multiply\m 60 \advance\m \time
  \ifnum \n>12 \advance\n -12 \fi
   \number\n.\twodigits\m~\ampm\time}
\def\ampm#1{\ifnum #1< 720 am\else pm\fi}
\def\twodigits#1{\ifnum #1<10 0\fi \number#1}

\begin{document}

%TITLE AND AUTHOR
\title{A Note on the Bayesian Approach to Sliding Window Detector Development}
\author{Graham  V. Weinberg  \\ (Draft created at \clock)\\
%Graham.Weinberg@dsto.defence.gov.au
 }
\maketitle

% The paper headers
\markboth{A Note on the Bayesian Approach  \today}%
{}

\begin{abstract}
Recently a Bayesian methodology has been introduced, enabling the construction of sliding window detectors with the constant false alarm rate property.
The approach introduces a Bayesian predictive inference approach, where under the assumption of no target, a predictive density of the cell under test, conditioned on the clutter range profile, is produced. The probability of false alarm can then be produced by integrating this density. As a result of this, for a given clutter model, the Bayesian constant false alarm rate detector is produced. This note outlines how this approach can be extended, to allow the construction of alternative Bayesian decision rules, based upon more useful measures of the clutter level.
\end{abstract}


\section{Introduction}
This note outlines how the Bayesian approach of \cite{weinberg18} can be extended, so that decision rules can be produced with the constant false alarm rate (CFAR) property. In the latter, in the context of Pareto Type II clutter, it is shown how one can produce a CFAR detector, based upon the idea of constructing the density of the cell under test (CUT), conditioned on the clutter range profile (CRP) and under the assumption that the return consists of clutter alone.
The probability of false alarm (Pfa) is then the integral of this density, from the threshold multiplier $\tau$ to $\infty$. Based upon this expression, one can then define the decision rule. It will be explained here that this technique can be modified very easily to produce CFAR detectors in any context, where the measurement of clutter is expressed by any appropriate statistic.

This note assumes that the reader is familiar with the problem of constructing non-coherent detection processes with the CFAR property, as fomulated in \cite{finn}. Useful guides on this approach include \cite{minkler90} and \cite{weinbergbook}.

\section{Preliminaries}
Suppose that the statistic of the CUT is $Z_0$ and that the CRP is modelled by the statistics $Z_1, Z_2, \ldots, Z_N$. It is assumed that all these statistics are independent and that they have the same common distribution function, including $Z_0$ in the absence of a target in the CUT.
Let $H_0$ to be the hypothesis that the CUT does not contain a target, and $H_1$ the alternative hypothesis that the CUT contains a target embedded within clutter. Then the binary test can be specified in the form 
\begin{equation}
Z_0 \hyptest \tau g(Z_1, Z_2, \ldots, Z_N), \label{primarytest}
\end{equation}
where the notation employed in \eqref{primarytest} means that $H_0$ is rejected in the case where $Z_0$ exceeds $\tau g(Z_1, Z_2, \ldots, Z_N)$.
The threshold multiplier $\tau$ is used so that the detection process \eqref{primarytest} can have its Pfa controlled adaptively.  The corresponding Pfa is given by the expression
\begin{equation}
\pfa = \prob(Z_0 > \tau g(Z_1, Z_2, \ldots, Z_N)| H_0), \label{primarytestpfa}
\end{equation}
where $\prob$ denotes probability. If the expression \eqref{primarytestpfa} is such that the threshold multiplier $\tau$ can be set independently of the clutter power, then the test \eqref{primarytest} will be referred to as a sliding window detector with the CFAR property. Since clutter power is a function of the clutter model's parameters, it is sufficient to show that $\tau$ does not depend on unknown clutter parameters to ascertain that \eqref{primarytest} is CFAR.
The corresponding probability of detection (Pd) is given by
\begin{equation}
\pd = \prob(Z_0 > \tau g(Z_1, Z_2, \ldots, Z_N)| H_1), \label{primarytestpd}
\end{equation}
and requires knowledge of the target model in order to extract a detection probability.

\section{Bayesian Approach}

In the Bayesian approach one constructs the Bayesian predictive distribution of the CUT, given the CRP, namely \\$Z_0 | Z_1, Z_2, \ldots, Z_N$, under the null hypothesis $H_0$. This is still under the usual independence assumptions. The difference is that underlying clutter parameters in the model are viewed as random variables, and as a result of this the CUT and CRP are coupled and thus statistically dependent. Once the density of the predictive  distribution has been constructed, it is possible to determine the threshold multiplier $\tau$ through the expression for the Pfa given by
\begin{eqnarray}
\pfa &=& \prob(Z_0 > \tau | Z_1=z_1, \ldots, Z_N=z_N) 
\int_{\tau}^\infty f_{Z_0|Z_1,\ldots, Z_N}(z_0 | z_1, \ldots, z_N) dz_0 \label{pfaexpBayes}
\end{eqnarray}
where $f_{Z_0|Z_1,\ldots, Z_N}(z_0 | z_1, \ldots, z_N)$ is the density of the predictive distribution. The Bayesian test is to reject $H_0$ if 
$Z_0 | Z_1=z_1, \ldots, Z_N=z_N > \tau$. The Bayesian predictive density can be derived from the likelihood function and the assumed prior distribution for the unknown parameters.

Consider the case of a one-parameter clutter model, with unknown parameter $\lambda$. Then the Bayesian predictive density is
\begin{eqnarray}
f_{Z_0 | Z_1, \ldots, Z_N}(z_0 | z_1, z_2, \ldots, z_N) 
&=& {\int_0^\infty}{f_{Z_0 | \Lambda} (z_0 | \lambda)} f_{\Lambda | Z_1, \ldots, Z_N}
(\lambda | z_1, z_2, \ldots, z_N) f_{\Lambda}(\lambda) d\lambda, 
\label{preddens}
\end{eqnarray}
where $\Lambda$ is the random variable modelling the unknown distributional parameter, $f_{Z_0 | \Lambda} (z_0 | \lambda)$ is the density of the CUT, conditioned on $\Lambda$, $f_{\Lambda | Z_1, \ldots, Z_N}(\lambda | z_1, z_2, \ldots, z_N)$ is the density of $\Lambda$, conditioned on the CRP and 
$f_{\Lambda}(\lambda)$ is the prior distribution for $\Lambda$. 

In \cite{weinberg18} the Pareto Type II case is examined, so that the clutter model has two unknown parameters. For a two parameter clutter model, the Bayesian predictive density is
\begin{eqnarray}
 f_{Z_0 | Z_1, Z_2, \ldots, Z_N}(z_0 | z_1, z_2, \ldots, z_N) =
 \int_0^\infty \int_0^\infty f_{Z_0| \A, \B}(z_0 | \alpha, \beta) 
 f_{\A, \B | Z_1, Z_2, \ldots, Z_N}(\alpha, \beta | z_1, z_2, \ldots, z_N) 
 f_{(\A, \B)}(\alpha, \beta)  d\alpha d\beta, \label{predBayes}
\end{eqnarray}
where $\A$ and $\B$ are the distributions of the clutter parameters, $f_{Z_0| \A, \B}(z_0 | \alpha, \beta)$ is the density of the CUT conditioned on the clutter parameters, $f_{\A, \B|Z_1, Z_2, \ldots, Z_N}(\alpha, \beta | z_1, z_2, \ldots, z_N)$ is the posterior density of the parameters $\A$ and $\B$ conditioned on the CRP, and $f_{(\A, \B)}(\alpha, \beta)$  is the prior distributions of the clutter parameters.

Expression \eqref{preddens} can be applied in the Exponentially distributed clutter case, with Jeffreys prior used as the non-informative prior. In this case it is not difficult to show that the Bayesian detector is the well-known cell-averaging CFAR. It is useful to observe that instead of constructing the predictive density of the CUT, given the CRP, one can instead base the approach instead on a function of the CRP.

Suppose that $T = T(Z_1, Z_2, \ldots, Z_N)$ is a statistic of the CRP. Then the Bayesian predictive density is
\begin{eqnarray}
f_{Z_0 | T}(z_0 | t) 
&=& {\int_0^\infty}{f_{Z_0 | \Lambda} (z_0 | \lambda)} f_{\Lambda | T}
(\lambda | t) f_{\Lambda}(\lambda) d\lambda, \nonumber\\
\label{preddens2}
\end{eqnarray}
where the conditional densities are self-explanatory. Then the corresponding Pfa is 
\begin{eqnarray}
\pfa &=& \prob(Z_0 > \tau | T=t) 
= \int_{\tau}^\infty f_{Z_0|T}(z_0 | t) dz_0. \label{pfaexpBayes3}
\end{eqnarray}
One can then construct CFAR detectors based upon \eqref{pfaexpBayes3}. This technique can also be applied to \eqref{predBayes} to produce a counterpart for two parameter clutter models. The hope is that this approach can be used to produce CFAR decision rules in the Pareto Type II case, to extend the work in \cite{weinberg18} and \cite{weinberg17}, where CFAR detectors, based upon order statistics and trimmed means, could be produced.

\section{A Simple Example}
Consider the case where the statistics in the CRP are modelled by Exponentially distributed variables, with parameter $\lambda$. The density of the Jeffreys prior can be shown to be proportional to the reciprocal of $\lambda$, so that $f_\Lambda(\lambda) = \frac{1}{\lambda}$. Then since the density of the $k$th order statistic (OS) of the CRP is
\begin{equation}
f_{Z_{(k)}}(t) = \lambda k {N \choose k} \left( 1 - e^{-\lambda t}\right)^{k-1} e^{-\lambda t(N-k+1)} \label{exposdens}
\end{equation}
it follows that 
\begin{equation}
\int_0^\infty f_{Z_{(k)}}(t) \frac{d\lambda}{\lambda} = t^{-1}, \label{der1}
\end{equation}
where the integral has been evaluated with a change of variables $\phi = e^{-\lambda t}$, and with appeal to the definition of the Beta function.
Consequently it follows that
\begin{equation}
f_{\Lambda | Z_{(k)}}(\lambda | t) = \lambda t k {N \choose k} \left( 1 - e^{-\lambda t}\right)^{k-1} e^{-\lambda t(N-k+1)}. \label{der2}
\end{equation}
Clearly it follows that
\begin{equation}
f_{Z_0 | \Lambda} (z_0 | \lambda) = \lambda e^{-\lambda z_0}. \label{der3}
\end{equation}
Hence applying \eqref{der2} and \eqref{der3}, together with the Jeffreys prior, to \eqref{preddens} one arrives at the predictive density
\begin{equation}
f_{Z_0| Z_{(k)} = t}(z_0) = kt{ N \choose k} \int_0^\infty \lambda \left( 1 - e^{-\lambda t}\right)^{k-1}  e^{-\lambda \left[ z_0 + t(N-k+1)\right]} d\lambda.
\label{der4}
\end{equation}
An application of the Binomial Theorem shows that
\begin{equation}
\left( 1 - e^{-\lambda t}\right)^{k-1} = \sum_{i=0}^{k-1} {k-1 \choose i} (-1)^i e^{-\lambda t i}. \label{der5}
\end{equation}
Applying \eqref{der5} to \eqref{der4} results in 
\begin{equation}
f_{Z_0| Z_{(k)} = t}(z_0) = kt{ N \choose k} \sum_{i=0}^{k-1} {k-1 \choose i} (-1)^i 
\int_0^\infty \lambda e^{-\lambda \left[ z_0 + t(N-k+1 + i)\right]} d\lambda. \label{der6}
\end{equation}
In order to evaluate the integral in \eqref{der6}, apply a transformation $\psi = \lambda \left[ z_0 + t(N-k+1 + i)\right]$ and with an appeal to the Gamma function it follows that \eqref{der6} reduces to 
\begin{equation}
f_{Z_0| Z_{(k)} = t}(z_0) = kt{ N \choose k} \sum_{i=0}^{k-1} {k-1 \choose i} (-1)^i \left[ z_0 + t(N-k+1 + i)\right]^{-2}. \label{der7}
\end{equation}
Finally, the Pfa can be produced by an application of \eqref{der7} to \eqref{pfaexpBayes}. Hence
\begin{equation}
\pfa(\tau) = kt{ N \choose k} \sum_{i=0}^{k-1} {k-1 \choose i} (-1)^i \left[ \tau + t(N-k+1 + i)\right]^{-1}. \label{der7}
\end{equation}
Hence the decision rule is to reject $H_0$ if $Z_0 > \tau$, where for a given Pfa $\tau$ is determined through \eqref{der7}.

Note that in the case where $k=1$ expression \eqref{der7} reduces to
\begin{equation}
\pfa(\tau) = Nt \left[\tau + Nt\right]^{-1},
\end{equation}
from which one can deduce that
\begin{equation}
\tau = t\left[N(\pfa^{-1} - 1)\right].
\end{equation}
Combining these results it follows in this simpler case that the decision rule is
\begin{equation}
Z_0 \hyptest N(\pfa^{-1}-1) Z_{(1)}
\end{equation}
which is the well-known CFAR decision rule based upon the CRP minimum.

It appears difficult to show that this result extends to the general case in \eqref{der7}. Recall that it is sufficient to compare $\pfa(z_0)$ with the design Pfa, and to reject $H_0$ if the former is smaller than the latter. This eliminates the need to invert \eqref{der7} numerically.

\end{document}